\documentclass[aps,prb,reprint,superscriptaddress,showpacs,showkeys,preprintnumbers,longbibliography]{revtex4-2}
\usepackage{graphicx}
\usepackage{bm}
\usepackage{amsmath}
\usepackage{amssymb}
\usepackage{subfigure}
\usepackage{float}
\usepackage{multirow}
\usepackage{booktabs}
\usepackage{varioref}
\usepackage{xr-hyper}
\usepackage{hyperref}
\usepackage{xcolor}
\usepackage{nicefrac}
\usepackage{xfrac}
\usepackage{hyperref}
\hypersetup{colorlinks,linkcolor=blue,urlcolor=blue,citecolor=blue}
\usepackage{ulem}
\usepackage{siunitx}
\usepackage{graphicx}
\usepackage{dcolumn}
\usepackage{braket}
\usepackage{wasysym}
\usepackage{textcomp}

\DeclareUnicodeCharacter{FF09}{fi}

\begin{document}

\title{Two-Dimensional Phase-Fluctuating Superconductivity in Bulk-Crystalline NdO$_{0.5}$F$_{0.5}$BiS$_2$}

\author{C. S. Chen}
\affiliation{State Key Laboratory of Surface Physics and Department of Physics, Fudan University, Shanghai 200433, China}
\affiliation{Physik-Institut, Universität Zürich, Winterthurerstrasse 190, CH-8057 Zürich, Switzerland}

\author{J. Küspert}
\affiliation{Physik-Institut, Universität Zürich, Winterthurerstrasse 190, CH-8057 Zürich, Switzerland}

\author{I. Biało}
\affiliation{Physik-Institut, Universität Zürich, Winterthurerstrasse 190, CH-8057 Zürich, Switzerland}
\affiliation{AGH University of Krakow, Faculty of Physics and Applied Computer Science, 30-059 Krakow, Poland}

\author{J. Mueller}
\affiliation{Physik-Institut, Universität Zürich, Winterthurerstrasse 190, CH-8057 Zürich, Switzerland}

\author{K. W. Chen}
\affiliation{State Key Laboratory of Surface Physics and Department of Physics, Fudan University, Shanghai 200433, China}

\author{M. Y. Zou}
\affiliation{State Key Laboratory of Surface Physics and Department of Physics, Fudan University, Shanghai 200433, China}

\author{D. G. Mazzone}
\affiliation{Laboratory for Neutron Scattering and Imaging, Paul Scherrer Institut, CH-5232 Villigen PSI, Switzerland}

\author{D. Bucher}
\affiliation{Physik-Institut, Universität Zürich, Winterthurerstrasse 190, CH-8057 Zürich, Switzerland}

\author{K. Tanaka}
\affiliation{Physik-Institut, Universität Zürich, Winterthurerstrasse 190, CH-8057 Zürich, Switzerland}
\affiliation{Department of Physics and Engineering Physics, University of Saskatchewan, 116 Science Place, Saskatoon, Saskatchewan, Canada S7N 5E2}

\author{O.~Ivashko}
\affiliation{Deutsches Elektronen-Synchrotron DESY, Notkestra{\ss}e 85, 22607 Hamburg, Germany.}

\author{M.~v.~Zimmermann}
\affiliation{Deutsches Elektronen-Synchrotron DESY, Notkestra{\ss}e 85, 22607 Hamburg, Germany.}

\author{Qisi Wang}
\affiliation{Department of Physics, The Chinese
University of Hong Kong, Shatin, Hong Kong, China}
\affiliation{Physik-Institut, Universität Zürich, Winterthurerstrasse 190, CH-8057 Zürich, Switzerland}

\author{Lei Shu}
\altaffiliation[Corresponding author: ]{leishu@fudan.edu.cn}
\affiliation{State Key Laboratory of Surface Physics and Department of Physics, Fudan University, Shanghai 200433, China}

\author{J. Chang}

\affiliation{Physik-Institut, Universität Zürich, Winterthurerstrasse 190, CH-8057 Zürich, Switzerland}

\begin{abstract}
 We present a combined growth and transport study of superconducting single-crystalline NdO$_{0.5}$F$_{0.5}$BiS$_2$. Evidence of two-dimensional superconductivity with significant phase fluctuations of preformed Cooper pairs preceding the superconducting transition is reported.
 This result is based on three key observations. (1) The resistive superconducting transition temperature $T_c$ (defined by resistivity $\rho \rightarrow 0$) increases with increasing disorder. (2) As $T\rightarrow T_c$, the conductivity diverges significantly faster than what is expected from Gaussian fluctuations in two and three dimensions. (3) Non-Ohmic resistance behavior is observed in the superconducting state.
 Altogether, our observations are consistent with a temperature regime of phase-fluctuating superconductivity. The crystal structure with magnetic ordering tendencies in the NdO$_{0.5}$F$_{0.5}$ layers and (super)conductivity in the BiS$_2$ layers is likely responsible for the two-dimensional phase fluctuations. As such, NdO$_{0.5}$F$_{0.5}$BiS$_2$ falls into the class of unconventional ``laminar" bulk superconductors that include cuprate materials and 4Hb-TaS$_2$.

\end{abstract}

\maketitle
\section{Introduction}
Conventional superconductivity is well described by the Bardeen-Cooper-Schrieffer~\cite{BCS} (BCS) theory or its strong-coupling extensions~\cite{Marsiglio2020}. The superconducting condensate constitutes a macroscopic wave function,
$\Psi=\Delta\exp(i\phi)$ with a pairing amplitude $\Delta$ and phase $\phi$. Pairing of Fermi-liquid quasiparticles~\cite{FermiL_1} and phase coherence emerge simultaneously below the critical temperature $T_c$. Phase stiffness is particularly pronounced in the limit where the Fermi energy is much larger than the pairing amplitude. BCS superconductors have no nodes in their energy gap and are typically insensitive to nonmagnetic impurities~\cite{Anderson-1959}.

Unconventional superconductivity in its broadest sense refers to the superconducting behavior that departs from the conventional BCS theory. In the dirty limit towards the superconductor-insulator transition as due to disorder or lowering of the dimensionality, even conventional $s$-wave superconductors exhibit a pseudogap at temperatures much higher than $T_c$ (see, e.g., Ref.~\onlinecite{Chand2012} and references therein). This originates in the presence of superconducting islands that fail to achieve global phase-coherence across the system~\cite{disorder-granular,Ghosal,Ghosal2,Dubi2007}.
In very disordered NbN~\cite{MondalPRL2011,Chand2012} and TiN~\cite{Sacp2010} thin films, for example, phase-fluctuating Cooper pairs exist prior to the superconducting transition, resulting in superconducting correlations present well above $T_c$.
Other examples are high-temperature cuprate~\cite{LeeRMP2007,Proust2019} and iron-based~\cite{Stewart2011,Hosono2015} superconductors, where unconventional superconductivity arises from pairing of non-Fermi-liquid quasiparticles~\cite{Daou2008,Cooper2009,Shibauchi2014,YFeGe2014}. Cuprates~\cite{Monthoux2007,Hashimoto2014} and some iron pnictides~\cite{Thomale2011,KFeAs2010,CsFeAs-PhysRevB} exhibit nodal superconductivity and are sensitive to nonmagnetic impurities~\cite{BernhardPRL1996,Yang2013,Li2015}, while cuprate superconductors are intrinsically disordered~\cite{Cren2001,Pan2001}.

The design principles of unconventional superconductivity remain to be an active field of research.
Confining materials in two dimensions is a common route to explore unconventional superconductivity~\cite{Saito2016}. However, in bulk crystals, it is challenging to completely decouple superconductivity along one direction. Even very tetragonal crystal structure can host finite interlayer Josephson coupling~\cite{Schafgans2010}.

\begin{figure*}[t]
\centering
\begin{minipage}[c]{0.999\textwidth}
  \centering
{\includegraphics[width=\textwidth]{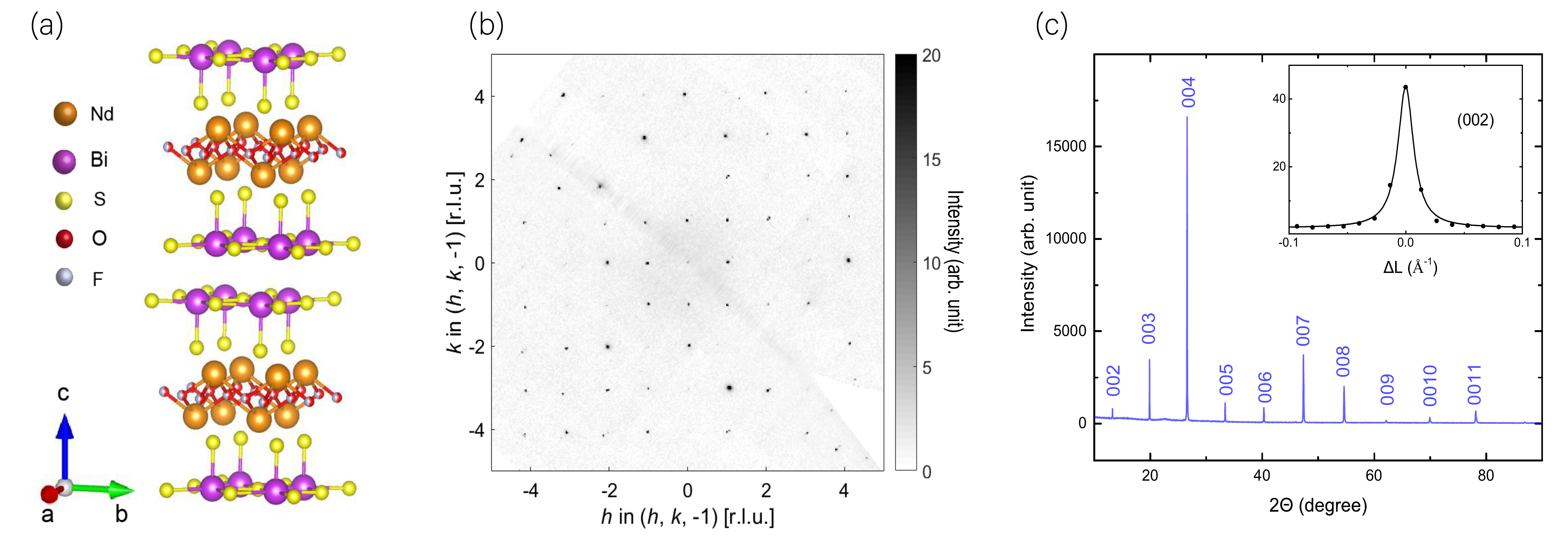}}
\end{minipage}
   \caption{Crystal structure and x-ray diffraction (XRD) results of NdO$_{0.5}$F$_{0.5}$BiS$_2$. (a) Crystal structure of NdO$_{0.5}$F$_{0.5}$BiS$_2$. (b) Representative high-energy (100 keV) XRD mapping of the Bragg reflections in the $(h,k,-1)$ plane. (c) XRD pattern of a NdO$_{0.5}$F$_{0.5}$BiS$_2$ single crystal measured with copper $K_\alpha$ x-rays. The $\theta$-$2\theta$ scan shows that the surface normal of the crystal is along the $(0,0,l)$ direction. The inset displays the line cut of the $(0,0,2)$ Bragg peak from the high-energy XRD data.}
  \label{1}
\end{figure*}

Here, we provide an improved growth procedure for NdO$_{0.5}$F$_{0.5}$BiS$_2$ leading to large high-quality single crystals. The observed paraconductivity exhibits strong deviation from the Gaussian fluctuation theory. This, combined with the observation of non-Ohmic $I-V$ characteristics and a strong disorder dependence of the superconducting transition temperature, provides evidence consistent with two-dimensional phase-fluctuating superconductivity in NdO$_{0.5}$F$_{0.5}$BiS$_2$. This dimensional reduction is likely linked to the magnetic ordering tendency of the NdO$_{0.5}$F$_{0.5}$ layers that in turn decouple the superconducting BiS$_2$ layers.

\section{Methods}

High-quality single crystals of NdO$_{0.5}$F$_{0.5}$BiS$_2$ were grown using CsCl/KCl flux~\cite{SampleGrow1}. The starting materials Nd, Bi, Nd$_2$O$_3$, NdF$_3$, Bi$_2$O$_3$, Bi$_2$S$_3$, and S were mixed in a nominal stoichiometric ratio, and the molar ratio of flux CsCl/KCl was CsCl : KCl = 5 : 3. Weighing and grinding of the raw materials were carried out in an argon atmosphere. The starting materials (0.8 g) and flux (5 g) were mixed and sealed in a high vacuum quartz tube. The inner surface of the quartz tube was coated with a carbon film to stop the flux from corroding the quartz tube.
A sealed quartz tube was heated to 800 $^{\circ}$C,  for 10 h, before cooled to 600 $^{\circ}$C at a rate of 0.5 $^{\circ}$C/h. Finally, we furnace-cooled the sample to room temperature. By removing residual flux with distilled water, we obtained high-quality single crystals. Thickness and lateral size of the crystals are, respectively, 10-100 $\mu$m and $\sim6$~mm.
The volume of our crystals is therefore 3-5 times larger than previously reported~\cite{BiS2_1,BiS2_2}.

We performed high-energy (100 keV) x-ray diffraction experiments on our single crystal of NdO$_{0.5}$F$_{0.5}$BiS$_2$ at the P21.1 beamline at PETRA III (DESY), and the single crystal Cu $K_\alpha$ (8.04 keV) x-ray diffraction was performed on Bruker D8 advance XRD spectrometer, which gives us robust evidence for high sample quality. Resistivity measurements were carried out on Quantum Design (QD) physical property measurement system (PPMS) with a constant DC current of 1~mA. Voltage-current characteristics were measured in a commercial PPMS. We used a Keithley-6220 precision current source to supply the current and the corresponding voltage was measured using Keithley 2182 nanovoltmeters equipped with preamplifiers~\cite{DestrazPRB2017,HomeBuilt2,HomeBuilt3}.

\begin{figure*}[t]
\centering
\begin{minipage}[c]{0.999\textwidth}
  \centering
{\includegraphics[width=\textwidth]{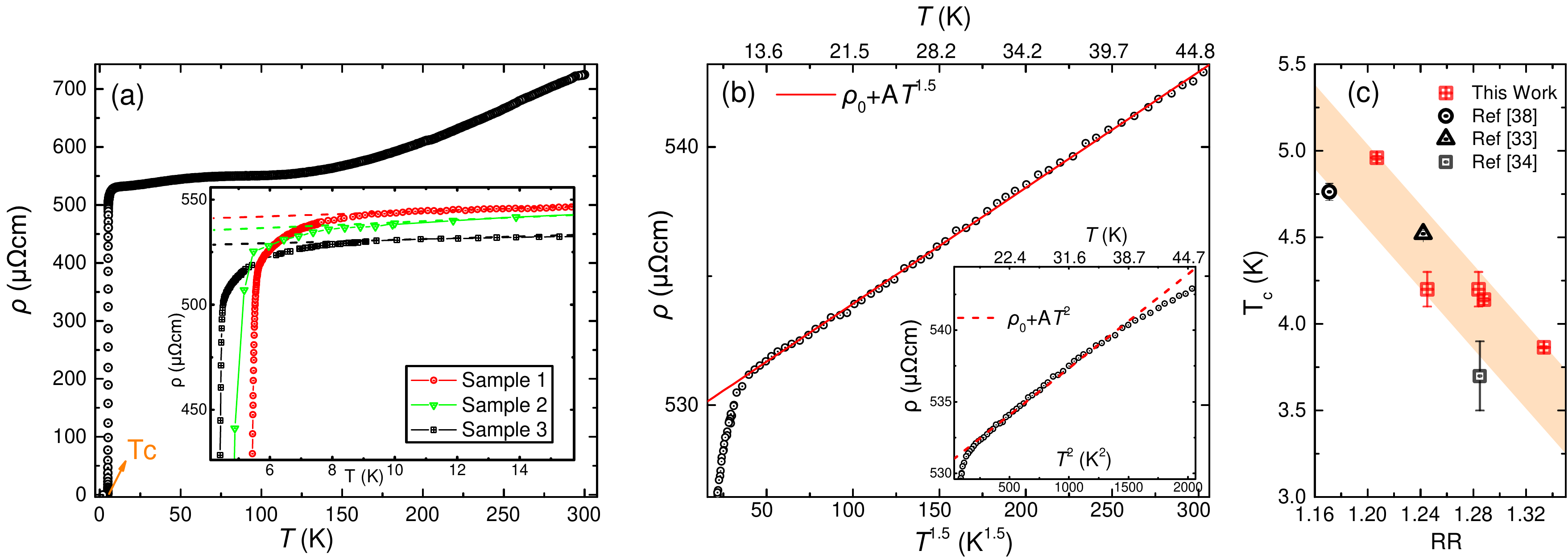}}
\end{minipage}
   \caption{In-plane resistivity $\rho_{ab}$ of NdO$_{0.5}$F$_{0.5}$BiS$_2$ versus temperature. (a) Temperature dependence of $\rho_{ab}$ with the transition temperature $T_c$ indicated.
   Inset shows a resistivity versus temperature for three different samples. Dashed lines indicate extrapolations to estimate the residual resistivity $\rho_0$.
   (b) Low-temperature resistivity plotted as a function of $T^{1.5}$.
   Inset shows the same data plotted versus $T^2$. (c) Superconducting transition temperature versus the resistivity ratio defined by $\rho(300~\textrm{K})/\rho(140~\textrm{K})$. Red points denote data from this work and black points represent literature values as indicated~\cite{BiS2_1,BiS2_2,RN1}. }
  \label{2}
\end{figure*}

\section{Results}
The layered P4/nmm structure of NdO$_{0.5}$F$_{0.5}$BiS$_2$ (space group \#129)  is shown in Fig.~\ref{1}(a). The structure is composed of alternately stacked superconducting BiS$_2$ bilayers and magnetic NdO$_{0.5}$F$_{0.5}$ layers~\cite{Review-BiS2}.
High-energy (100 keV) x-ray diffraction recorded at room temperature reveals excellent (single) crystallinity. Bragg features within the $(h,k,-1)$ scattering plane are shown in Fig.~\ref{1}(b). In Fig.~\ref{1}(c), we show Cu $K_\alpha$ (8.04 keV) x-ray diffraction data along the reciprocal out-of-plane $(0,0,\ell)$ direction.
Also here high crystallinity and the absence of impurity phases are observed.
The inset of Fig.~\ref{1}(c) displays the $(0,0,2)$ Bragg reflection measure with 100 keV photons. The Bragg peak width corresponds to an out-of-plane correlation length $\xi_c\simeq 125$~\AA, indicating excellent stacking order.

The temperature dependence of the in-plane resistivity $\rho_{ab}$, shown in Fig.~\ref{2}(a), is consistent with previous reports~\cite{BiS2_1,BiS2_2}. Even within the same growth batch, slightly different residual resistivity $\rho_0$ values are found. Motivated by the resistivity plateau in the temperature range $80 - 140$~K, we define the resistivity ratio as RR = $\rho(300~\textrm{K})/\rho(140~\textrm{K})$. Generally, we find $\rho_0\sim 500 - 550$~$\mu\Omega$cm and RR=$1 - 1.4$ across our grown samples. The low RR values suggest that NdO$_{0.5}$F$_{0.5}$BiS$_2$ is a disordered superconductor. The superconducting transition temperature (defined by the temperature below which the resistance is indistinguishable from zero)  varies in the range $T_c=3.5-5$~K. In fact, $T_c$ and RR appear to anti-correlate -- see Fig.~\ref{2}(c), where data from Refs.~\onlinecite{BiS2_1,BiS2_2,RN1} are also shown. Samples with lower RR and higher $\rho_0$ values have a higher transition temperature.

In what follows, we describe results on one of our samples.
In Fig.~2(b), the low-temperature resistivity is plotted as a function of $T^{1.5}$ and $T^{2}$ (see inset). We find that the $T^{1.5}$ dependence describes the resistivity over a wider temperature range. A fit to $\rho_n = \rho_0 + A  T^{1.5}$ yields $\rho_0 = \rm{529}$~$\rm{\mu \Omega cm}$ and $A= 0.0449~\rm{\mu \Omega cm K^{-1.5}}$.

\begin{figure*}[thb]
\centering
\begin{minipage}[c]{1\textwidth}
  \centering
{\includegraphics[width=\textwidth]{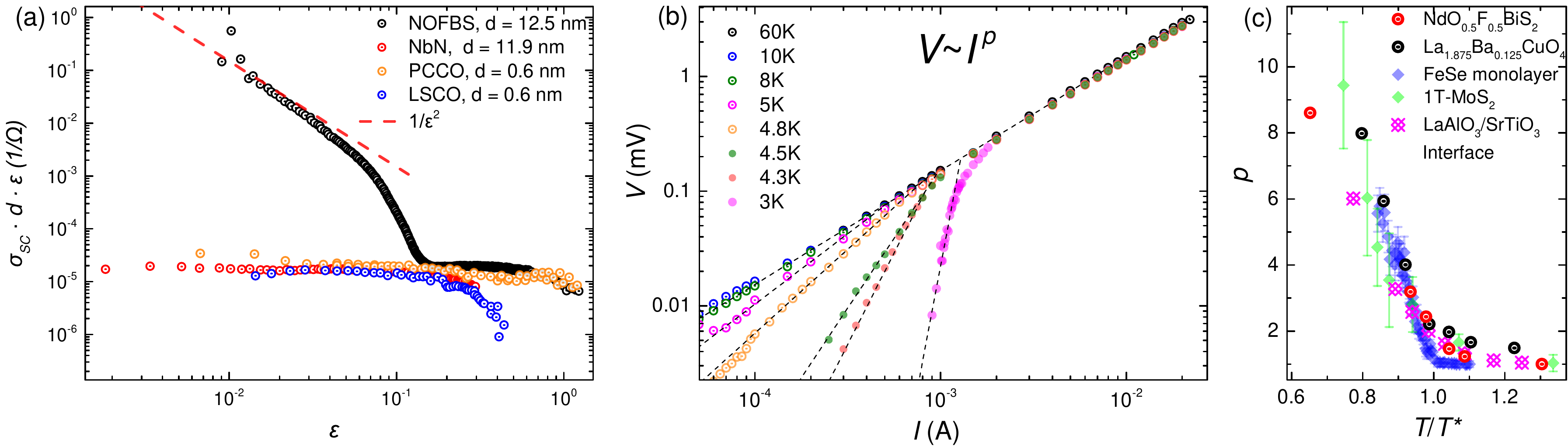}}
\end{minipage}
   \caption{
   Unconventional paraconductivity and voltage-current characteristics.
   (a) compares paraconductivity plotted as $\sigma_{sc}\,d\,\epsilon$ (see Eq.~(1)) versus $\epsilon=(T-T_c)/T_c$ for NdO$_{0.5}$F$_{0.5}$BiS$_2$, ${\rm Pr}_{2-x}{\rm Ce}_x{\rm CuO_4}$~\cite{PCCO},  ${\rm La}_{2-x}{\rm Sr}_x{\rm CuO_4}$~\cite{LSCO}, and NbN~\cite{DestrazPRB2017}.
   (b) V-I curves in log-log scale for temperatures as indicated. Dotted lines are power-law V $\sim$ I$^p$ scaling with $p$ adjusted to fit the data.
   (c) Temperature dependence of $p$ for NdO$_{0.5}$F$_{0.5}$BiS$_2$ (red points), compared with results on La$_{1.875}$Ba$_{0.125}$CuO$_4$ 	(black points)~\cite{2D1}, monolayer FeSe (blue points)~\cite{FeSe_ML}, 1T-MoS$_2$ (green points)~\cite{MoS2}, and LaAlO$_3$/SrTiO$_3$ interface (magenta points)~\cite{LAO}. }
  \label{3}
\end{figure*}

Next, we turn to observations of paraconductivity. Our analysis assumes the applicability of the Matthiessen rule~\cite{Matthiessen}. That is, $\sigma=\sigma_{sc} + \sigma_{qp}$, where for zero magnetic field $\sigma_{qp}=1/\rho_n$ refers to the normal-state quasiparticle transport and $\sigma_{sc}$  is the conductivity from short-lived superconducting Cooper pairs or phase fluctuating superconductivity. With $\sigma=1/\rho$, we infer the conductivity from superconducting fluctuations: $\sigma _{sc} = \sigma - \sigma _{qp} $. In Fig.~\ref{3}(a), we compare $\sigma_{sc}$ for NbN~\cite{DestrazPRB2017}, ${\rm Pr}_{2-x}{\rm Ce}_x{\rm CuO_4}$ (PCCO)~\cite{PCCO},  ${\rm La}_{2-x}{\rm Sr}_x{\rm CuO_4}$ (LSCO)~\cite{LSCO} and NdO$_{0.5}$F$_{0.5}$BiS$_2$ (NOFBS) as a function of distance $\epsilon=(T-T_c)/T_c$ to the superconducting transition temperature $T_c$. For NbN, PCCO and LSCO, $\sigma_{sc}$ scales with $\epsilon^{-1}$ for $\epsilon\rightarrow 0$ as expected from standard Gaussian fluctuations in two-dimensional systems. In fact, for NbN the expected $\sigma _{sc} = e^2/(16 \hbar d \epsilon)$ is observed over more than one order of magnitude in $\epsilon$~\cite{1985-NbN}. For NdO$_{0.5}$F$_{0.5}$BiS$_2$, the $\sigma _{sc}=e^2/(16 \hbar d\epsilon)$ scaling is found for intermediate values of $\epsilon$. However, as $\epsilon\rightarrow 0$, strong deviation from the $\epsilon^{-1}$ scaling is observed with much faster divergence.

This unconventional behavior of the superconducting fluctuations led us to investigate the $I-V$ characteristics.
Figure~\ref{3}(b) shows $I-V$ curves in a double logarithmic scale for various temperatures as indicated. For $T>T^{*}\approx 5$~K, the standard Ohmic ($I\propto V$) behavior is found. Inside the superconducting state, however, we find deviation from the Ohmic behavior below a critical current $I_c \sim 1$~mA. For $T<T^{*}$ and $I<I_c$, the $I-V$ curves can be described by a power-law dependence, $I\propto V^{-p}$. With decreasing temperatures below $T_c$, the exponent $p$ increases -- see Fig.~\ref{3}(c).
Such temperature dependence of $p$ is observed for two-dimensional superconductivity hosted by, for example, monolayer FeSe~\cite{FeSe_ML} or the interface between SrTiO$_3$ and LaAlO$_3$~\cite{LAO}.

\section{Discussion}
The $T^{1.5}$ dependence of the normal-state resistivity observed in NdO$_{0.5}$F$_{0.5}$BiS$_2$ could originate from proximity to a magnetic quantum critical point. It is known that CeO$_{0.5}$F$_{0.5}$BiS$_2$ orders ferromagnetically and exhibits a lower superconducting transition temperature~\cite{CeBiS2_1,CeBiS2_2,CeBiS2_3,CeBiS2_4}. It is therefore not inconceivable that  NdO$_{0.5}$F$_{0.5}$BiS$_2$ hosts critical spin fluctuations that generate the non-Fermi-liquid behavior~\cite{LohneysenRMP2007}. It is also not uncommon to find superconductivity around such a magnetic quantum critical point~\cite{ScallapinoRMP2012,Shibauchi2014}. In addition, there are a few examples of unconventional superconductivity emerging from $\rho\sim AT^{1.5}$ non-Fermi liquids such as in KFe$_2$As$_2$ in the dirty limit~\cite{KFeAs2010}, CsFe$_2$As$_2$~\cite{CsFeAs-PhysRevB} and YFe$_2$Ge$_2$~\cite{YFeGe2014}. In these three materials~\cite{KFeAs2010,CsFeAs-PhysRevB,YFeGe2014} as well as our NdO$_{0.5}$F$_{0.5}$BiS$_2$ samples, we find no apparent correlation between the scattering coefficient $A$ and $T_c$. This is in contrast to electron-doped cuprates where a positive correlation between $A$ in $\rho\sim AT$ and $T_c$ has been found~\cite{A_Tc_2010}.

The crystal structure with BiS$_2$ bilayers separated by NdF$_{0.5}$O$_{0.5}$ layers makes a potential host for two-dimensional electronic orders.
A large resistivity anisotropy $\rho_c/\rho_{ab}\approx 1500$ has been reported for Pr$_{1.05}$O$_{0.82}$F$_{0.18}$Bi$_{1.03}$S$_2$~\cite{Nagao2015}, suggesting two-dimensional electronic structure~\cite{HorioPRL2018}. ARPES experiments on
NdO$_{0.5}$F$_{0.5}$BiS$_2$ has demonstrated that the band structure is highly two-dimensional~\cite{YePRB2014}. Electronic two-dimensionality can be enhanced further when neighboring layers host different orders. In 4Hb-TaS$_2$~\cite{Ribak2020} superconductivity is sandwiched by Mott insulating layers. Another example is  La$_{1.875}$Ba$_{0.125}$CuO$_4$~\cite{2D1}, where alternating stripe order is believed to quench the $c$-axis Josephson coupling.
Superconductivity in NdO$_{0.5}$F$_{0.5}$BiS$_2$ is likely confined within the BiS$_2$ layers and the ground state involves magnetism in the NdF$_{0.5}$O$_{0.5}$ layers. In fact, a density functional theory study of NdO$_{0.5}$F$_{0.5}$BiS$_2$ claims two possible magnetic ground states at low temperatures~\cite{DFT-BiS}.
Therefore, NdO$_{0.5}$F$_{0.5}$BiS$_2$ is expected to host highly two-dimensional superconductivity.

 Upon approaching the superconducting transition temperature, the coherence length $\xi$ diverges and substantially exceeds the out-of-plane lattice parameter and the electronic mean free path $\ell$.   As discussed above, NdO$_{0.5}$F$_{0.5}$BiS$_2$ may belong to the class of highly resistive two-dimensional superconductors. Such superconductors are expected to display Gaussian fluctuations. In this case, the conductivity from short-lived Cooper pairs is expected to show a power-law divergence:
\begin{equation}
\it \sigma_{sc} \simeq \frac{e^2}{16\hbar d}\frac{1}{\epsilon} ,
\label{AL}
\end{equation}
where $\hbar$ and $e$ are, respectively, the reduced Planck constant and the elementary charge~\cite{PourretNatPhys2006}. The length scale that confines superconductivity in two dimensions is labeled $d$.
For film systems, $d$ is typically defined as the film thickness. Two-dimensional superconductivity emerges when the out-of-plane superconducting coherence length $\xi_{sc}^c$ exceeds $d$. In this limit, Gaussian fluctuations provide the conductivity channel expressed in Eq.~(\ref{AL}), namely, $\sigma_{sc}d\epsilon \sim e^2/(\rm{16}\hbar)$, a constant as a function of $\epsilon$. Plotted in Fig.~\ref{3}(a) are $\sigma_{sc}d\epsilon$ from data on a NbN film~\cite{DestrazPRB2017} with film thickness $d\simeq 100$~\AA, ${\rm Pr}_{2-x}{\rm Ce}_x{\rm CuO_4}$~\cite{PCCO}, and ${\rm La}_{2-x}{\rm Sr}_x{\rm CuO_4}$~\cite{LSCO}, which are all independent of $\epsilon$. To reach this numerical consistency for cuprates (films or crystals), $d$ is made comparable to the layer spacing $c/2\sim 6$~\AA. The c-axis coherence length $\xi_{sc}^c$ is typically much shorter than the ab-plane coherence length $\xi_{sc}^{ab}$ in cuprates, yet $c/2\lesssim \xi_{sc}^c$ ($\sim 8\,$\AA\ and $\sim 7\,$\AA, respectively, for PCCO and LSCO~\cite{Wu2014}).
In contrast, NbN presents strongly coupled $s$-wave superconductivity with an isotropic coherence length, which is larger than the film thickness $d$.

As can be seen in Fig.~3(a), $\sigma_{sc}$ obtained on our bulk crystals of NdO$_{0.5}$F$_{0.5}$BiS$_2$ shows strikingly different dependence on $\epsilon$, when the out-of-plane lattice correlation length $\xi_c$ is taken as the confining length scale, i.e., $d= \xi_c$ = 125~\AA. It follows standard two-dimensional Gaussian fluctuations for $\epsilon > 0.2$, whereas significant deviation from $\sigma_{sc}\sim \epsilon^{-1}$ is observed for $\epsilon <0.2$. In NdO$_{0.5}$F$_{0.5}$BiS$_2$, $\sigma_{sc}$ grows rapidly as $T$ approaches $T_c$ and eventually an approximately $\sigma_{sc}\sim \epsilon^{-3}$ power-law growth emerges in the $\epsilon\rightarrow0$ limit. This strongly suggests the existence of non-Gaussian fluctuations. Phase fluctuations from preformed Cooper pairs are a possible source for this sudden rise of $\sigma_{sc}$. This implies that NdO$_{0.5}$F$_{0.5}$BiS$_2$ displays both amplitude- and phase-fluctuating superconductivity above $T_c$. As the contribution of phase fluctuations to the conductivity decays faster as $\epsilon$ increases, Gaussian fluctuations dominate for $\epsilon>0.2$. Conversely, phase fluctuations are the dominant contribution as $\epsilon \rightarrow 0$. This corroborates our observation of non-Ohmic $V-I$ behavior that is commonly observed in phase-fluctuating two-dimensional superconductors.

Superconductivity is often sensitive to disorder. For example, in monolayer FeSe two-dimensional superconductivity emerges only in the clean limit~\cite{FeSe_ML}. On the contrary, in NbN films the phase fluctuating regime is reached in the limit where disorder localizes the electronic wave functions~\cite{MondalPRL2011}. Moreover, in thin films of several soft metals such as Al and Sn, larger $T_c$ has been observed for higher sheet resistance~\cite{Disorder-Tc-01,Disorder-Tc-02,Disorder-Tc-04,Disorder-Tc-05,yeh2023doubling}. The same phenomenon is also reported in bulk aluminum-copper alloy~\cite{Disorder-Tc-03}. Also in NdO$_{0.5}$F$_{0.5}$BiS$_2$, higher residual resistivity seems to favor unconventional superconductivity. The value of $\rho_0\sim 500 - 550$~$\mu\Omega$cm of our samples is smaller but the same order of magnitude as those found in La$_{2-x}$Ba$_x$CuO$_4$~\cite{Moodenbaugh1988} and
underdoped Ba(Fe$_{1-x}$Co$_{x}$)$_2$As$_2$~\cite{Chu2010}, and an order of magnitude smaller than observed in underdoped cuprates~\cite{chang2012LSCO}. In NdO$_{0.5}$F$_{0.5}$BiS$_2$ as well as NbN and TaS$_2$~\cite{TaS2-disorder}, the large sheet resistance stems likely from chemical disorder. The corresponding localization of the electronic wave functions may affect the superconducting properties including $T_c$. We find in our samples that the smaller the RR and the larger the $\rho_0$, the higher the $T_c$. Mechanisms for enhancing $T_c$ by disorder in unconventional as well as conventional superconductors have been proposed~\cite{Bergmann-1973,disorder-theory-n1,disorder-theory-n2,Romer2018,Gastiasoro2018} and such enhancement has been observed in La$_{1.875}$Ba$_{0.125}$CuO$_4$~\cite{Leroux2019} and the simple metals mentioned above~\cite{Disorder-Tc-01,Disorder-Tc-02,Disorder-Tc-04,Disorder-Tc-05,yeh2023doubling,Disorder-Tc-03}.

It is also worth noting that phase fluctuations above $T_c$ are expected to occur in strongly coupled superconductors. Scanning tunnelling spectroscopy experiments indicate that $2\Delta/ (k_BT_c) = 16.8$ with $\Delta$ being the superconducting pairing amplitude for NdO$_{0.5}$F$_{0.5}$BiS$_2$~\cite{BiS2_2,Yazici2015}. This ratio is more than four times larger than expected from the weak-coupling BCS theory. Hence it is reasonable to assume a strong coupling scenario for NdO$_{0.5}$F$_{0.5}$BiS$_2$. All these evidences combined point to strong-coupling superconductivity with unusually large fluctuations of preformed Cooper pairs in bulk crystalline  NdO$_{0.5}$F$_{0.5}$BiS$_2$.

\section{Conclusion}
In summary, we have successfully grown large high-quality single crystals of NdO$_{0.5}$F$_{0.5}$BiS$_2$. The single crystal quality has been demonstrated through x-ray diffraction measurements. Resistivity scales with $T^{1.5}$ before entering the regime of superconducting fluctuations. The observations of non-Ohmic $I-V$ characteristics, non-Gaussian superconducting fluctuations, and disorder dependence of the superconducting transition provide evidence of a two-dimensional phase-fluctuating regime above the transition temperature. This dimensional reduction is likely due to magnetic ordering tendencies in the NdO$_{0.5}$F$_{0.5}$ layers that effectively decouple the superconducting BiS$_2$ layers.\\

\begin{acknowledgments}
J.K. and J.C. acknowledge support by the Swiss National Science Foundation (Projects No. 200021-188564).  I.B. acknowledges support from the Swiss Confederation through the Government Excellence Scholarship. C.S.C., K.W.C., M.Y.Z., and L.S. acknowledge support by the National Key Research and Development Program of China (Project No. 2022YFA1402203), the National Natural Science Foundations of China (Project No. 12174065), and C.S.C. also acknowledges support by China Scholarship Council. K.T. acknowledges support by the Pauli Center for Theoretical Studies. Q.W. is supported by the Research Grants Council of Hong Kong (ECS No. 24306223), and the CUHK Direct Grant (No. 4053613). Parts of this research were carried out at the PETRA III beamline P21.1 at DESY, a member of the Helmholtz Association (HGF). The research leading to this result has been supported by the project CALIPSOplus under the Grant Agreement 730872 from the EU Framework Programme for Research and Innovation HORIZON 2020.
\end{acknowledgments}
\vfill

\end{document}